\documentclass[10pt,aps,prb,twocolumn,showpacs,floatfix,superscriptaddress]{revtex4-1}
\usepackage{amsmath,amsbsy,graphics,epsfig,mathrsfs}
\usepackage{bm}
\usepackage[]{color}

\newcommand{\tx}{\text}
\newcommand{\ti}{\textit}

\newcommand{\nn}{\nonumber}
\newcommand{\pat}{\partial}
\newcommand{\para}{\parallel}

\newcommand{\alp}{\alpha}

\newcommand{\Dlt}{\Delta}

\newcommand{\eps}{\epsilon}

\newcommand{\Gm}{\Gamma}
\newcommand{\lam}{\lambda}

\newcommand{\etal}{{\em et al.,~}}
\newcommand{\ie}{{i.e.,~}}
\usepackage{wrapfig}

\begin{document}
\title{Angular dependence and symmetry of Rashba spin torque in ferromagnetic heterostructures}
\author{Christian Ortiz Pauyac}
\affiliation{King Abdullah University of Science and Technology (KAUST),
Physical Science and Engineering Division, Thuwal 23955-6900, Saudi Arabia}
\author{Xuhui Wang}
\affiliation{King Abdullah University of Science and Technology (KAUST),
Physical Science and Engineering Division, Thuwal 23955-6900, Saudi Arabia}
\author{Mairbek Chshiev}
\affiliation{SPINTEC, UMR CEA/CNRS/UJF-Grenoble 1/Grenoble-INP, INAC, Grenoble, F-38054, France}
\author{Aurelien Manchon}
\email[]{aurelien.manchon@kaust.edu.sa}
\affiliation{King Abdullah University of Science and Technology (KAUST),
Physical Science and Engineering Division, Thuwal 23955-6900, Saudi Arabia}
\date{\today}

\begin{abstract}
In a ferromagnetic heterostructure, the interplay between a Rashba
spin-orbit coupling and an exchange field gives rise to a
current-driven spin torque. In a realistic device setup, we
investigate the Rashba spin torque in the diffusive regime and
report two major findings: (i) a nonvanishing torque exists at the
edges of the device even when the magnetization and effective
Rashba field are aligned; (ii) anisotropic spin relaxation rates
driven by the Rashba spin-orbit coupling assign the spin torque a
general expression ${\bm T}=T^y_{\para}(\theta){\bm
m}\times(\hat{\bm y}\times{\bm m})+T^y_{\bot}(\theta)\hat{\bm
y}\times{\bm m}+T^z_{\para}(\theta){\bm m}\times(\hat{\bm
z}\times{\bm m})+T^z_{\bot}(\theta)\hat{\bm z}\times{\bm m}$,
where the coefficients $T_{\para,\bot}^{y,z}$ depend on the
magnetization direction. Our results agree with recent
experiments.
\end{abstract}
\maketitle

The concept of current-driven spin-orbit torque in ultrathin
ferromagnetic heterostructures\cite{man-zhan-aRtorque-2008,
expmiron1} and diluted magnetic semiconductors\cite{dms} is
attracting much attention for providing an efficient magnetization
switch mechanism using just one ferromagnet. In contrast to the
conventional spin-transfer torque that demands a spin-polarized
current generated by a reference ferromagnet
(polarizer),\cite{slow-stt-1996,refstt} the spin-orbit torque
accomplishes magnetization switching by transferring angular
momentum between the spin and orbital degrees of freedom through a
spin-orbit coupling. In ferromagnetic heterostructures typically
made of magnetic trilayers comprising an ultrathin ferromagnetic
film sandwiched between a noble metal and an insulator,
experiments and theories have uncovered a spin-orbit torque of the
form\cite{expmiron1,wang-manchon-2011,theoryall}
\begin{equation}
{\bm T} = T_\para{\bm
m}\times(\hat{\bm y}\times{\bm m})+T_{\perp}\hat{\bm y}\times{\bm m},
\label{eq:torque-general}
\end{equation}
where ${\bm m}$ is the magnetization direction and $\hat{\bm y}$
is the directional unit vector (see Fig. 1). Two components in
Eq.(\ref{eq:torque-general}) are usually referred to as {\em
in-plane} ($T_{\para}$) and {\em perpendicular} ($T_\bot$)
torques.

The current understanding of the origin of spin-orbit torque in
ferromagnetic heterostructures combines spin-Hall
effect\cite{she-all} and band structure-induced effects. In the
former, a spin current is generated by the spin-Hall effect in the
noble metal layer and injected into the ferromagnet to produce a
torque.\cite{lius,haney} In the latter, symmetry breaking across
the interface between the noble metal and the ferromagnet induces
a spin splitting in the band structure, leading to a nonvanishing
current-induced spin-orbit field. One version of this band
structure-induced spin-orbit torque is the so-called Rashba
torque\cite{man-zhan-aRtorque-2008} where an electric field ${\bm
E}$ embedded across the interface produces a nonequilibrium Rashba
field $\bm{B}_R\propto\hat{\bm{z}}\times\bm{E}$. Recent works have
emphasized the complexity of these torques as functions of
materials parameters.\cite{kim} Of most interest to the present
study, it has been shown that the spin-orbit torque possesses a
complex angular dependence that is not captured by the earlier
models based on either spin-Hall effect or Rashba
torques.\cite{angulardep}

\begin{figure}[ht]
\centering
\includegraphics[trim = 33mm 65mm 10mm 60mm, clip, scale=0.53]{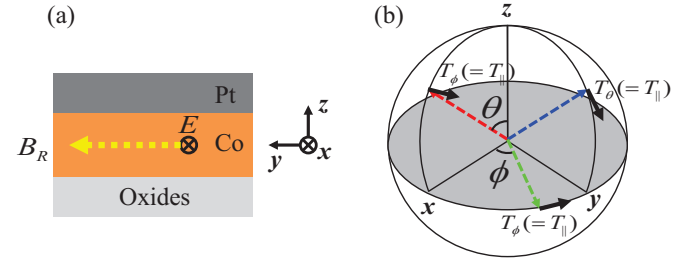}
\caption{(Color online)(a) Schematic view the device cross
section. A charge current is flowing along the $x$ direction,
generating an effective Rashba field $\bm{B}_R$ (dotted yellow
line) that is pointing to $y$. $L$ is the size of the
lateral dimensions. (b) Spin torque components $\bm{T_{\phi}}$ and
$\bm{T_{\theta}}$ in a spherical coordinate, with $\theta$ and $\phi$ being the azimuthal and the in-plane angle, respectively. The coordinates dashed
lines show the magnetization directions considered in this
Letter.} \label{fig:representation}
\end{figure}

In this Letter, we employ a set of coupled spin-charge diffusion
equations\cite{wang-manchon-2011} to characterize the Rashba spin
torque in a square-shaped device with a width (and
length) $L$, see Fig.\ref{fig:representation}. We highlight two
major findings. First, even when the magnetization $\bm{m}$ is
aligned along the effective spin-orbit field (along the $\hat{\bm y}$
direction), the spin-orbit torque does \ti{not} vanish at the
edges of the device. Second, the torque amplitudes vary with the
magnetization direction (a behavior to be referred as \ti{angular
dependence}) and are size-dependent. In the weak (spin-orbit)
coupling regime, the angular dependence vanishes as the device
size increases; whereas in a strong coupling regime, the angular
dependence is insensitive to the device size.

A schematic view of the device is shown in Fig.
\ref{fig:representation}(a). The inversion asymmetry across the
interfaces generates a Rashba type spin-orbit coupling from the
potential gradient along the $\hat{\bm{z}}$ direction. In the
\ti{quasi-two-dimensional} system considered here, the diffusive
dynamics of nonequilibrium spin density $\bm{S}$ and charge
density $n$ are described by \cite{wang-manchon-2011,arxiv_rashba}
\begin{align}
\frac{\pat n}{\pat t} =& D\nabla^{2}n+ B {\bm\nabla}_{xy}\cdot{\bm
S}+\Gamma {\bm\nabla}_{xy}\cdot {\bm m} n +R
{\bm\nabla}_{xy}\cdot \bm{m}(\bm{S}\cdot \bm{m}),
\label{eq:charge-diffusion}\\
\frac{\pat \bm{S}}{\pat t} =& D \nabla^{2}{\bm S} -\frac{\bm
S}{\tau_{sf}} -\frac{\bm S +S_z\bm{\hat z}}{\tau_{\tx{DP}}}
-\frac{\bm{m}\times(\bm{S}\times \bm{m})}{T_{xc}}+B {\bm\nabla}_{xy} n \nn\\
&-\Dlt_{xc}\bm{S}\times \bm{m}
+2 C {\bm\nabla}_{xy}\times{\bm S}
+2 R ( \bm{m}\cdot{\bm\nabla}_{xy} n) \bm{ m}\nn\\
&+\Gamma \left[ \bm{m}\times({\bm\nabla}_{xy}\times{\bm S})
+{\bm\nabla}_{xy}\times({\bm m}\times{\bm S})\right].
\label{eq:spin-dynamics}
\end{align}
The parameters are: $\hbar=1$, $C =\alp v_Fk_{F}\tau$,
$\Gamma=\alpha\Dlt_{xc}v_F k_F\tau^2/2$, $R
=\alp\Dlt_{xc}^{2}\tau^{2}/2$, and $B =2\alpha^3 k_F^2\tau^2$.
$\alp$ denotes the Rashba spin-orbit coupling and ${\bm\nabla}_{xy} =
\bm{\hat z}\times\bm{\nabla}$. The momentum relaxation time is
$\tau$ and spin relaxation time due to magnetic impurities is
given phenomenologically by $\tau_{sf}$, whereas
$\tau_{\tx{DP}}=1/2\alp^2k_F^2\tau$ is the D'yakonov-Perel
relaxation time.\cite{dp-she-1971} $k_F(v_{F})$ is the Fermi
momentum (velocity) while $D=v_{F}^2\tau/2$ is the diffusion
constant. Briefly, $\Dlt_{xc}$- and $C$-terms describe spin
precession around the exchange field and $\bm{B}_{R}$,
respectively. $B$-term couples spin and charge degrees of freedom,
leading to the electrical spin generation and spin-Hall effect.
$\Gm$-term provides a higher order correction to the precessional
motion described by first two terms. $R$-term contributes to a
magnetization renormalization. $T_{xc}=1/\Dlt_{xc}^2\tau$ is the
transverse spin dephasing time in the limit of a weak
ferromagnet.\cite{dephasing} As a boundary condition suggested by
experiments,\cite{kato-science} the nonequilibrium spin density is
required to vanish at the transverse edges along the $y$
direction.

\begin{figure}[ht]
\centering
\includegraphics[scale=0.25]{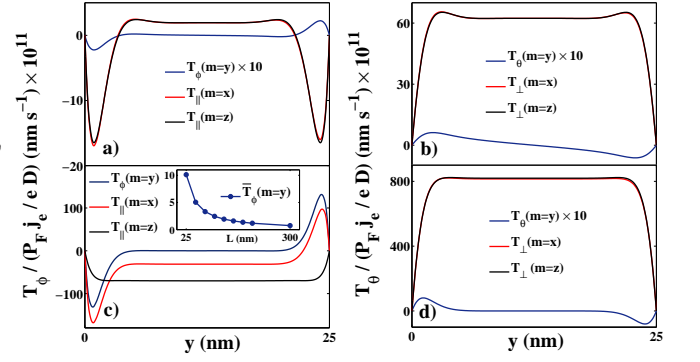}
\caption{(Color online) Spatial distribution of the Rashba torque
along the $y$ axis, for a device $L=25$~nm. The $yz-$plane is
located at the center of the device, \ie $x=12.5$~nm. Panels (a)
and (c) refer to the \ti{in-plane} component (here $T_{\para}
\equiv T_{\phi}$). Panels (b) and (d) are for the
\ti{out-of-plane} component ($T_{\perp} \equiv T_{\theta}$).
Panels (a) and (b): weak Rashba regime ($\alpha = 0.001~\tx{eV
nm}$, $\Delta_{xc} = 0.1~\tx{eV}$). Panels (c) and (d): strong
coupling ($\alpha = 0.05~\tx{eV nm}$, $\Delta_{xc} =
0.01~\tx{eV}$). The inset in panel (c) displays:
$\overline{T}_{\phi(m=y)}=\int_{y=L/2}^{y=L}T_{\phi(m=y)}dy/L$ for
different widths. In panels (a),(b) and (d),  ${T}_{\phi(m=y)}$ is
multiplied by a factor 10. The Fermi energy is $E_{F}=0.7$~eV,
$k_{F}=4.3~\tx{nm}^{-1}$ and $v_{F}=5\times 10^{14}~\tx{nm
s}^{-1}$. $\tau=10^{-15}~\tx{s}$ and $\tau_{sf}=10^{-12}~\tx{s}$.}
\label{fig:torque-cross-section}
\end{figure}

In a recent work, it was shown that reducing the width of the
magnetic film dramatically modifies the relative {\em magnitude}
of the in-plane and perpendicular torques in the weak Rashba
limit. \cite{wang-manchon-2011} Here, we further argue that
reducing the size of the device results in changes in the {\em
symmetry} of the torque. To support our argument, we plot in Fig.
\ref{fig:torque-cross-section} the spatial distribution of the
spin torque density (along the $y$ axis in the $yz-$plane) for
various magnetization directions in both weak ($\Delta_{xc}\gg
\alpha k_F$) and strong ($\Delta_{xc}\ll \alpha k_F$) spin-orbit coupling
regimes. In Fig. \ref{fig:representation}(b), the spin torque
density is expressed in spherical coordinates, $\bm{T} =
T_{\phi}\hat{\bm e}_\phi + T_{\theta}\hat{\bm e}_\theta$, which is
more general than Eq. (\ref{eq:torque-general}). On the right
column of Fig. \ref{fig:torque-cross-section}, $T_{\theta}$ is robust in the bulk,
resulting from a robust nonequilibrium spin density ($S_y$) driven
by the spin-galvanic effect discussed by Edelstein.\cite{edel}
This effect disappears towards the boundaries, as imposed by the
boundary conditions.\cite{kato-science}

An important feature in Fig. \ref{fig:torque-cross-section}
appears to be the nonvanishing spin torque at the edges even when
the magnetization $\bm{m}$ is parallel to $\bm{B}_{R}$
(deep blue curves). In general, as the angle between the exchange
field $\bm{m}$ and $\bm{B}_{R}$ closes, the spin torque amplitude
decreases. For a strong spin-orbit coupling, the spin-Hall effect
drives oppositely polarized spin densities $S_z$ accumulating at
opposite edges.\cite{she-all} In our finite-size device, within
the distance of spin-flip relaxation length from the edges, the
spin density $S_{z}$ distributed along the $y$ direction generates
a nonvanishing local spin torque (at the edges) when $\bm{m}$ is
parallel to $\bm{B}_{R}$. As the spin-orbit coupling weakens,
torques at $\bm{m}\para\bm{B}_{R}$ driven by the spin-Hall effect
become negligible, see Fig.\ref{fig:torque-cross-section}(a).
Meanwhile, as the sample size increases $L\gg \lam_{sf}$, the
expression
$\overline{T}_{\phi(m=y)}=(1/L)\int_{y=L/2}^{y=L}T_{\phi(m=y)}dy$
exponentially decreases, see inset in Fig.
\ref{fig:torque-cross-section}(c). Another important feature is
the inhomogeneous profile of $T_\phi$, which
is driven by the competing effects of spin precession around the
total field and spin Hall-effect:  the spin-Hall effect
(precession around the total field) is dominant in the strong
(weak) Rashba regime. When $\bm{m}=\bm{\hat z}$, spin density
$S_z$ does not contribute to the spin-Hall induced torque and the
in-plane torque becomes homogeneous, as confirmed by the solid
black line in Fig.\ref{fig:torque-cross-section}(c). In a finite
size device, these effects contribute to the angular dependence
discussed in the following.

First, in an infinite system with a weak Rashba spin-orbit
coupling, once the anisotropy in spin relaxation rates due to the
D'yakonov-Perel mechanism is quenched, Eq.
(\ref{eq:spin-dynamics}) gives rise to a torque described by Eq.
(\ref{eq:torque-general}).\cite{wang-manchon-2011} However, as
long as the spin relaxation rate is not isotropic, the torque
assumes a more complex angular dependence. In an infinite system, setting 
 ${\bm\nabla}_{xy} = \bm{\hat z}\times eE\partial_\eps\hat{\bm{x}}$, Eq.
(\ref{eq:spin-dynamics}) reduces to
\begin{align}
\Delta_{xc}{\bm S}\times{\bm m}&+\frac{1}{T_{xc}}{\bm
m} \times({\bm S}\times{\bm m})\nn\\
&+\frac{1}{\tau_{xy}}S_x{\bm x}+\frac{1}{\tau_{xy}}S_y\hat{\bm y}+\frac{1}{\tau_{z}}S_z\hat{\bm z}={\bm X},
\label{eq:spin-accu-analytical}
\end{align}
where the last three terms on the left-hand side subscribe to both
the D'yakonov-Perel mechanism and spin relaxation induced by
random magnetic impurities:
$\tau_{xy}^{-1}=\tau_{DP}^{-1}+\tau^{-1}_{sf}$ and
$\tau_z^{-1}=2\tau_{DP}^{-1}+\tau^{-1}_{sf}$. 
On the right-hand side, the source
term reads
\begin{align}
{\bm X}\equiv \frac{nE}{\epsilon_F}(B\hat{{\bm
y}} + 2 C P \hat{{\bm y}}\times{\bm m}+\Gamma P{\bm m}\times(\hat{{\bm y}}\times{\bm m})),
\end{align}
where $P$ is the spin polarization at Fermi level ($\epsilon_F$).
Equation (\ref{eq:spin-accu-analytical}) can be analytically
solved in spherical coordinates using $\bm{T} = T_{\phi}\hat{\bm
e}_\phi + T_{\theta}\hat{\bm e}_\theta$. In the strong coupling
limit ($B\gg C,\Gamma$), the spin torque becomes
\begin{align}
\bm T = \frac{BnE}{\epsilon_F}\epsilon_\theta&[{\beta} \bm{\hat y}\times \bm m
+ (1+(\xi-{\beta})\xi)\bm{m}\times\bm{\hat y}\times\bm{m}\nn\\
&-{\chi}(1-\alp_\theta\cos^2\theta)(\bm{m}\cdot\hat{\bm{x}})\bm{m}\times\bm{\hat{z}}\times\bm{m}\nn\\
&-\alp_\theta m_z m_y(1+(\xi-\beta)\xi)\bm{m}\times\bm{\hat{z}}\times\bm{m}\nn\\
&+\chi(\xi-\beta)(1-\alp_\theta\cos^2\theta)(\bm{m}\cdot\hat{\bm{x}})\bm{\hat z}\times\bm{m}\nn\\
&-\beta\alp_\theta m_zm_y\bm{\hat z}\times\bm{m}],
\label{eq:torque-odd-even}
\end{align}
which comprises one of the major results of this Letter. For a
succinct discussion, the parameters in
Eq.(\ref{eq:torque-odd-even}) are defined in
Ref.\onlinecite{parameter-defs}.

Equation (\ref{eq:torque-odd-even}) contains both odd and even
components with respect to the inversion of magnetization
direction ($\bm{m}$), which agrees with the expressions proposed
by Garello \ti{et al} [see Eqs.(9) and (10) in Ref.
\onlinecite{angulardep}]. In particular, besides the regular {\em
in-plane} and {\em perpendicular} torques captured by Eq.
(\ref{eq:torque-general}), additional terms in the form of
$\hat{\bm z}\times{\bm m}$ and ${\bm m}\times(\hat{\bm
z}\times{\bm m})$ arise. The relative magnitude of the different
contributions depends on the materials parameters. Furthermore, it
is interesting to notice that such a complex angular dependence of
the torque is solely determined by the anisotropy in
spin-relaxation rates (times). By suppressing the anisotropy, \ie
$\tau_z=\tau_{xy}$ and $1/\tau_{-}=0$, the torque reduces to
\begin{align}
\bm{T} = \frac{BnE}{\eps_F}\left[\frac{{\beta}}{1+\xi^2} \hat{\bm{
y}}\times \bm{m} +
\left(1-\frac{{\beta}\xi}{1+\xi^2}\right)\bm{m}\times\bm{\hat
y}\times\bm{m}\right], \label{eq:torque-simple}
\end{align}
[see Eq.(\ref{eq:torque-general})] and the complex angular
dependence vanishes. In our model, this anisotropic spin
relaxation is determined by the D'yakonov-Perel mechanism arising
from scatterings in the presence of Rashba spin-orbit
coupling.\cite{dp-she-1971} We emphasize here that the above
analytical results are obtained in a sample of infinite size in
the strong Rashba coupling regime.

\begin{figure}[ht]
\centering
\includegraphics[width=3.5in]{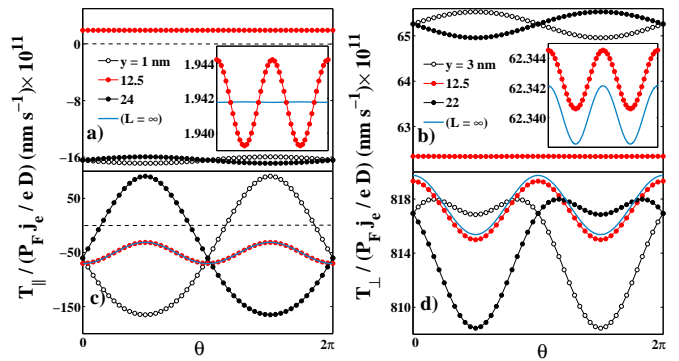}
\caption{(Color online) Angular dependence of spin torque as $\bm
m$ varies in the $xz-$plane in a system with $L= 25~\tx{nm}$.
Panels (a) and (c): in-plane torque; panels (b) and (d):
out-of-plane component. Panels (a) and (b): weak Rashba regime;
panels (c) and (d): strong Rashba regime. Solid red curves are
taken at the center of the device.
Empty and solid black dots are taken at positions near two
boundaries along the $y$ direction. Insets in panels (a),(b) give
a better picture of the solid red curves. Solid cyan lines refer
to the spin torque for $L=\infty$.}
\label{fig:angxz}
\end{figure}

In the following, we show that the angular dependence of the
Rashba torque shown in Eq. (\ref{eq:torque-odd-even}) also exists
in a device of \ti{finite} size. In addition, we also explain the
symmetry properties of spin torque at sample edges, as shown in
Fig. \ref{fig:angxz}. We analyze the angular dependence for
various $\bm{m}$ in the $xz-$plane at three particular locations
in the device, \ie in the center (at $x,y =12.5~\tx{nm}$) to
represent bulk values, and two other locations near the edges
along $y$ (at $x=12.5~\tx{nm}$). In what follows, the description
is taken considering $T_{\para}$ and $T_{\perp}$, defined in Eq.
(\ref{eq:torque-general}).

In the weak Rashba regime, when the magnetization is along
$\hat{\bm{z}}$, the spin density components that contribute to the
torque show a symmetric profile.\cite{wang-manchon-2011} As
$\bm{m}$ moves towards the $\hat{\bm{x}}$ axis, the spin density
generated perpendicular to $(\bm{m},\bm{B}_{R})$ points to the
$\hat{\bm{z}}$ direction. However, the faint presence of spin-Hall
effect still renders the total profile asymmetric, \ie one edge is
more negative than the other, in contrast to the case when $\bm{m}
= \hat{\bm{z}}$. Such an effect contributes to the angular
dependence at the edge of the device, as depicted by the open and
filled black dots in Fig.\ref{fig:angxz}(a)(b). In the strong
Rashba regime, the spin-Hall effect is dominating, producing a
more pronounced angular dependence, as shown by the open and
filled black dots in Fig.\ref{fig:angxz}(c)(d).

To illustrate the above effects when magnetization is in the $xz-$
plane, we study the angular dependence in the bulk for different
device sizes (in Fig. \ref{fig:angxz}, $L = 25~\tx{nm}$ and
$L=\infty$, only). In the strong Rashba regime, the spin
relaxation rate is dominated by D'yakonov-Perel term, the angular
dependence is pronounced and it does not change as the device size
increases, which shall eventually approach the limit characterized by
Eq.(\ref{eq:torque-odd-even}). In contrast, in the weak Rashba
regime the relaxation rate is mostly isotropic, which results in a
weak angular dependence that vanishes as the size increases. These
numerical results are consistent with the argument that the
angular dependence of spin torque in the bulk is driven by the
anisotropy in spin relaxation rate. Furthermore, in a
finite system with isotropic spin relaxation rates, oscillations
may arise due to edge effects diffusing towards the center and
such a phenomenon is better seen in the weak Rashba regime [see
inset in Fig. \ref{fig:angxz}(a)].
\begin{figure}[ht]
\centering
\includegraphics[width=3.5in]{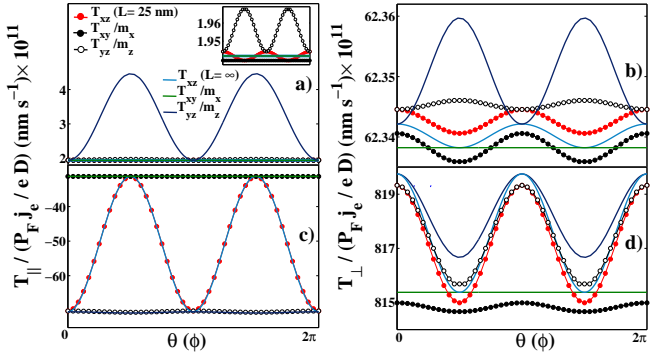}
\caption{(Color online) Angular dependence as $\bm m$ varies in
the \ti{xz}, \ti{xy} and \ti{yz} planes in the center of the
system for $L = 25~\tx{nm}$ and $L = \infty$. Panels (a) and (c):
in-plane torque; panels (b) and (d): out-of-plane component.
Panels (a) and (b) weak Rashba regime; panels (c) and (d) strong
Rashba regime. $T_{ij}$ refers to the spin torque in the $ij-$ plane.
The spin torque is divided by $m_x=\cos\phi$ ($m_z=\cos\theta$) in the $xy-$ ($yz-$) plane.}
\label{fig:lastxyyz}
\end{figure}
In Fig. \ref{fig:lastxyyz}, the spin torque density in the bulk is
plotted when $\bm{m}$ rotates in the $xz-$, $xy-$ and $yz-$
planes.  For the latter two planes, the angular dependence coming
from the cross product in Eq. \ref{eq:torque-general} is removed
(division by $\cos\phi(\cos\theta)$). Therefore, the oscillations of
the torque magnitude in these three cases can be fitted by
$K_1+K_2\sin^2\phi(\sin^2\theta)+K_3\sin^4\phi(\sin^4\theta)$, which is
consistent with Eq. (\ref{eq:torque-odd-even}). A prominent
feature is that the oscillations in the $xy$-plane vanish as the
device size increases; whereas in the $yz$-plane, the oscillations
persist even with an isotropic relaxation rate, which is due to
$\Gm$-term contributing to the angular dependence when $\bm{m}$ is
in the $yz$-plane, thus enhancing the spin torque amplitude in the
weak Rashba regime [see Fig. \ref{fig:lastxyyz}(a)(b)].

In conclusion, for a finite-size Rashba torque device, we have
shown that the spin torque is nonvanishing at the edges of the
sample even when the magnetization and the effective Rashba field
are parallel, as a result of the competition between spin-Hall
effect and the nonequilibrium spin density generated due to
anisotropic spin relaxation rates. Furthermore, the symmetry and
angular dependence of the spin torque are in general more
complicated that the conventional form assumed to date. In our calculations, the angular dependence is much larger for $T_{||}$ than for $T_{\perp}$. For a
sample of an infinite size, we have obtained an analytical
expression for the spin-orbit torque that shows both odd and even
components against magnetization inversion and agrees favorably
with the expression proposed based on experimental results. In
a view of increasing industrial and academic interests in the
field of spin-orbit torques, we expect that results presented in
this Letter shall not only provide a better understanding to the
key mechanisms behind the experimental observations but also shed
light on the design of realistic devices.

We thank I. M. Miron, K. Garello, K. -J. Lee, P. M. Haney and M.
D. Stiles for valuable discussions.


\begin{thebibliography}{999}
\bibitem{man-zhan-aRtorque-2008}
A. Manchon and S. Zhang, Phys. Rev. B {\bf 78}, 212405 (2008);
Phys. Rev. B {\bf 79}, 212405 (2009); I. Garate and A. H.
MacDonald, Phys. Rev. B {\bf 80}, 134403 (2009); A. Matos-Abiague
and R. L. Rodriguez-Suarez, Phys. Rev. B {\bf 80}, 094424 (2009).
\bibitem{expmiron1}
I. M. Miron \etal Nature Mate. {\bf 9}, 230 (2010); U. H. Pi \etal
Appl. Phys. Lett. {\bf 97}, 162507 (2010); I. M. Miron, \etal
Nature (London) {\bf 476}, 189 (2011).
\bibitem{dms}
A. Chernyshov, \etal Nature Phys. {\bf 5}, 656 (2009); D. Fang,
\etal Nature Nanotech. {\bf 6}, 413 (2011); K. M. D. Hals, A.
Brataas and Y. Tserkovnyak, Europhys. Lett. {\bf 90}, 47002 (2010).
\bibitem{slow-stt-1996}
J. C. Slonczewski, J. Magn. Magn. Mater. {\bf 159}, L1 (1996); L.
Berger, Phys. Rev. B {\bf 54}, 9353 (1996).
\bibitem{refstt}
D. C. Ralph and M. D. Stiles, J. Magn. Magn. Mater. {\bf 320},
1190 (2008), and references therein.
\bibitem{wang-manchon-2011}
X. Wang and A. Manchon, Phys. Rev. Lett. {\bf 108}, 117201 (2012).
\bibitem{theoryall}
K.-W. Kim \etal Phys. Rev. B {\bf 85} 180404(R) (2012); D. A.
Pesin and A. H. MacDonald, Phys. Rev. B {\bf 86} 014416 (2012); E.
van der Bijl and R. A. Duine, Phys. Rev. B {\bf 86}, 094406
(2012).
\bibitem{she-all}
J. E. Hirsch, Phys. Rev. Lett. 83, 1834 (1999); S. Zhang, Phys.
Rev. Lett. 85, 393 (2000).
\bibitem{lius}
L. Liu \etal Phys. Rev. Lett. {\bf 109}, 096602 (2012); L. Liu
\etal Science {\bf 336}, 555 (2012).
\bibitem{haney}
P.M. Haney \etal, arXiv:1301.4513
\bibitem{kim}
J. Kim \etal, Nature Materials {\bf12}, 240 (2013).
\bibitem{angulardep}
K. Garello, \etal arXiv:1301.3573
\bibitem{arxiv_rashba}
X. Wang, C. Ortiz Pauyac, and A. Manchon, arXiv:1206.6726, (2012).
\bibitem{dp-she-1971}
M. I. D'yakonov and V. I. Perel,
Sov. Phys. JETP Lett. {\bf 13}, 467 (1971).
\bibitem{dephasing}
C. Petitjean, D. Luc, and X. Waintal, Phys. Rev. Lett. {\bf 109}
117204 (2012); Y. Tserkovnyak, A. Brataas, and G. E. W. Bauer, J.
Magn. Magn. Mater. {\bf 320}, 1282 (2008).
\bibitem{kato-science}
Y. K. Kato, \etal Science {\bf 306}, 1910 (2004).
\bibitem{edel}
V. M. Edelstein, Solid State Commun. {\bf 73}, 233 (1991).
\bibitem{parameter-defs}
Here, we provide the parameters used in
Eq.(\ref{eq:torque-odd-even})
\begin{align}
\epsilon_\theta =& \left[1+\xi^2+\xi{\chi}\sin^2\theta
\left(1-\alp_{\theta}\cos^2\theta\right)\right]^{-1},\\
\alp_\theta =& \frac{\tau_{\para}(\theta)}{\tau_{-}},
~\frac{1}{\tau_{\para}(\theta)}=\frac{1}{\tau_{xy}}+\frac{\cos^2\theta}{\tau_{-}}
\end{align}
where $\xi=\frac{\tau_{\Dlt}}{\tau_{+}}$,
${\beta}=\frac{\tau_{\Dlt}}{\tau_{xy}}$,
${\chi}=\frac{\tau_{\Dlt}}{\tau_{-}}$,
$\tau_{\Dlt}=\frac{1}{\Dlt_{xc}}$,
$\frac{1}{\tau_{+}}=\frac{1}{T_{xc}}+\frac{1}{\tau_{xy}}$, and
$\frac{1}{\tau_{-}}=\frac{1}{\tau_{z}}-\frac{1}{\tau_{xy}}$.
\end{thebibliography}
\end{document}